\pdfoutput=1
\documentclass[12pt,preprint]{aastex}
\usepackage{amssymb,fancyhdr,lastpage,layout,bold-extra,floatrow,calc,wrapfig}
\usepackage{graphicx}
\usepackage{natbib}
\usepackage{aas_macros}
\usepackage[section] {placeins}
\usepackage{subfigure}
\usepackage{xspace}
\usepackage{float}
\usepackage{color}

\usepackage[scaled]{helvet}

\usepackage[T1]{fontenc}

\def\msun{{\rm\,M_\odot}}

\def\msun{{\rm\,M_\odot}} 

\def\zsun{{\rm\,Z_\odot}}

\newcommand{\kms}{\, {\rm km\, s}^{-1}}

\newcommand{\mpc}{\, {\rm Mpc}}
\newcommand{\kpc}{\, {\rm kpc}}
\newcommand{\hmpc}{\, h^{-1} \mpc}

\newcommand{\lya}{Ly$\alpha$ }

\newcommand{\hi}{\mbox{H\,{\scriptsize I}\ }}
\newcommand{\heii}{\mbox{He\,{\scriptsize II}\ }}

\def\h2{${\rm\,H_2}$}

\makeatletter 

\def\hi{\noindent \hangindent=2.5em}

\def\kpc{{\rm\,kpc}}

\def\mpc{{\rm\,Mpc}}

\def\kms{{\rm\,km/s}}
\def\msun{{\rm\,M_\odot}}

\def\vol#1  {{{#1}{\rm,}\ }}
\def\lya{{\rm Ly}\alpha}

\def\hi{{H\thinspace I}\ }

\def\heii{{He\thinspace II}\ }

\newcount\refno
\newcount\rfno
\def\eq{$^{\the\refno\ }$\advance\refno by 1}
\def\ad{\advance\rfno by 1}

\def\clock{\count0=\time \divide\count0 by 60
     \count1=\count0 \multiply\count1 by -60 \advance\count1 by \time
     \number\count0:\ifnum\count1<10{0\number\count1}\else\number\count1\fi}

\def\myputfigure#1#2#3#4#5%
{\hskip0.03\textwidth\vskip#5pt
\makebox[0pt]{\hskip#2in
\includegraphics[width=#3\textwidth]{#1}}\vskip#4pt\hfill}

\newcount\refno
\newcount\rfno
\def\eq{$^{\the\refno\ }$\advance\refno by 1}
\def\ad{\advance\rfno by 1}

\definecolor{burntorange}{rgb}{1,0.4,0.2}

\definecolor{burntorange}{rgb}{1,0.4,0.2}

\usepackage{color}

\begin{document}

\title{UV Absorption Line Ratios in Circumgalactic Medium at Low Redshift in Realistic Cosmological Hydrodynamic Simulations}

\author{Renyue Cen\altaffilmark{1} and Mohammadtaher Safarzadeh\altaffilmark{2}}

\footnotetext[1]{Princeton University Observatory, Princeton, NJ 08544;
 cen@astro.princeton.edu}

\footnotetext[2]{School of Earth and Space Exploration, Arizona State University, Tempe, AZ 85287-1404, USA;
Johns Hopkins University, Department of Physics and Astronomy, Baltimore, MD 21218, USA}

\begin{abstract} 
Utilizing high-resolution cosmological hydrodynamic simulations
we investigate various ultra-violet absorption lines 
in the circumgalactic medium of star forming galaxies at low redshift,
in hopes of checking and alleviating the claimed observational conundrum 
of the ratio of NV to OVI absorbers, among others.
We find a satisfactory agreement between simulations and extant observational data
with respect to the ratios of the following four line pairs examined,
NV/OVI, SiIV/OVI, NIII/OVI and NII/OVI.
For the pairs involving nitrogen lines, we examine two cases of nitrogen abundance,
one with constant N/O ratio and the other with varying N/O ratio,
with the latter motivated by theoretical considerations of two different synthetic sources of nitrogen
that is empirically verified independently.
Along a separate vector, for all line pairs, we examine two cases of radiation field,
one with the Haardt-Madau background radiation field and the other with an additional local radiation field sourced
by hot gas in the host galaxy. 
In all cases, two-sample Kolmogorov-Smirnov tests indicate excellent agreements.
We find that the apparent agreements between simulations and observations 
will be strongly tested, if the bulk of current upper limits of various line ratios are turned into actual detections.
We show that an increase in observational sensitivity by $0.2$ dex will already start to significantly 
constrain the models.
\end{abstract}

\section{Introduction}

The nature of halo gas (a.k.a. circumgalactic medium; CGM hereafter), on galactocentric distances of $10-500$ kpc,
is a problem of significant ongoing interest in galaxy formation.
Halo gas links galaxies from the intergalactic medium and 
is the conduit for exchange of matter energy density, momentum, angular momentum and metals
between star formation and active galactic nucleus induced outflow and gravitational inflow of gas.
Thus, understanding halo gas 
is imperative before a satisfactory theory of galaxy formation and evolution may be constructed.
There has been recent rapid advances on the observational front to address this important issue,
made possible primarily by HST observations 
\citep[e.g.,][]{2009Chen, 2011bProchaska, 2011bTumlinson, 2011Tripp, 2012Thom, 2013Werk, 2014Werk, 2014Peeples,2016Werk}.
In \citet[][]{2013Cen} we investigate the overall gas composition with respect to the density,
temperature and metallicity and find that,  
on average, for $>0.1L_*$ ({\color{red} red}, {\color{blue} blue}) galaxies
cold ($T<10^5$K) 
gas is the primary component in the inner regions,
with its mass comprising 50\% of all gas within $r=({\color{red}30},{\color{blue}150})$ kpc.
At $r>({\color{red}30},{\color{blue}200})$ kpc for ({\color{red}red}, {\color{blue}blue}) galaxies,
the hot ($T>10^7$K) component becomes the majority component.
The warm ($T=10^{5-7}$K) component is, on average, a perpetual minority in both red and blue galaxies,
with its contribution peaking at $\sim 30\%$ at $r=100-300$ kpc in blue galaxies
and never exceeding 5\% in red galaxies.
These findings are in agreement with recent observations in many aspects,
in particular with respect to the amount of warm gas in star forming galaxies
and the amount of cold gas in early type galaxies at low redshift,
both of which are physically intriguing and at first instant less than intuitive.

In light of a new observational development with respect to the NV to OVI absorption line ratio
and in particular the apparent need of seemingly complicated, perhaps contrived, models 
to explain the data,
we here perform a detailed analysis of our high resolution cosmological hydrodynamic simulations
to assess whether {\it ab initio} cosmological simulations are capable of accounting for this particular observation,
in the larger context of the success of the model able to match the overall composition of halo gas, 
among others. 
It is particularly relevant to note that the good agreement between our simulations and 
observations with respect OVI $\lambda \lambda$1032, 1038 absorption lines, presented earlier in \citet[][]{2012bCen},
suggests that the statistical description of the properties 
of the warm component in the simulations - mass, spatial distribution, density, temperature,
metallicity, and their environmental dependences - has now been firmly validated and provides
a critical anchor point for our model.
Consequently, this additional, independent analysis with respect to NV/OVI ratio and other ratios
becomes very powerful to further strengthen or falsify our model or our simulations.

Our findings here are both encouraging and intriguing.
If one uses a fixed, solar N/O ratio regardless of the O/H ratio, our model is acceptable,
with all 4 KS (Kolmogorov-Smirnov) test p-values greater than 0.28 for either 
Haardt-Madau \citep[][HM hereafter]{2012Haardt} 
or HM+local radiation field, where the local radiation field is due to hot gas in the host galaxy.
If one allows for a dependence of the N/O ratio on the O/H ratio,
both measured by independent observations and motivated by theoretical considerations of two 
different sources of N,
then our model is able to account for the observations highly successfully, 
with all KS test p-values exceeding 0.9.
We additionally examine the following absorption line column density ratios where comparisons to observations may be 
made in a reasonable statistical fashion, SiIV/OVI, NII/OVI and NIII/OVI,
and find that the ratios from our simulations are fully consistent with observations.
We also investigate the model where UV radiation from local shock heated gas in concerned galaxies 
are added to the HM background, which is found to also agree
with observations with comparable p-values for all line ratios examined.
However, these good agreements come about because observational data points are dominated
by upper and lower limits instead of actual detections.
We discuss how some moderate improvment in observational sensitivity 
may provide much stronger tests of models.

\section{Simulations}\label{sec: sims}

\subsection{Hydrocode and Simulation Parameters}

We perform cosmological simulations with the AMR Eulerian hydro code, Enzo \citep[][]{2014Bryan}.
We use the following cosmological parameters that are consistent with 
the WMAP7-normalized \citep[][]{2011Komatsu} LCDM model:
$\Omega_M=0.28$, $\Omega_b=0.046$, $\Omega_{\Lambda}=0.72$, $\sigma_8=0.82$,
$H_0=100 h {\rm km s}^{-1} {\rm Mpc}^{-1} = 70 {\rm km} s^{-1} {\rm Mpc}^{-1}$ and $n=0.96$.
These parameters are also consistent with the latest Planck results
\citep[][]{2014Planck},
if one adopts the Hubble constant that is the average between Planck value and those
derived based on SNe Ia and HST key program \citep[][]{2011Riess, 2012Freedman}.
We use the power spectrum transfer functions for cold dark matter particles and 
baryons using fitting formulae from \citet[][]{1999Eisenstein}.
We use the Enzo inits program to generate initial conditions.

First we ran a low resolution simulation with a periodic box of $120h^{-1}$Mpc on a side.
We identified two regions separately, one centered on
a cluster of mass of $\sim 2\times 10^{14}\msun$
and the other centered on a void region at $z=0$.
We then re-simulate each of the two regions separately with high resolution, but embedded
in the outer $120h^{-1}$Mpc box to properly take into account large-scale tidal field
and appropriate boundary conditions at the surface of the refined region.
We name the simulation centered on the cluster ``C" run
and the one centered on the void  ``V" run.
The refined region for ``C" run has a size of $21\times 24\times 20h^{-3}$Mpc$^3$
and that for ``V" run is $31\times 31\times 35h^{-3}$Mpc$^3$.
At their respective volumes, they represent $1.8\sigma$ and $-1.0\sigma$ fluctuations.
The root grid has a size of $128^3$ with $128^3$ dark matter particles.
The initial static grids in the two refined boxes
correspond to a $1024^3$ grid on the outer box.
The initial number of dark matter particles in the two refined boxes
correspond to $1024^3$ particles on the outer box.
This translates to initial condition in the refined region having a mean interparticle-separation of 
$117h^{-1}$ kpc comoving and dark matter particle mass of $1.07\times 10^8h^{-1}\msun$.
The refined region is surrounded by two layers (each of $\sim 1h^{-1}$Mpc) 
of buffer zones with 
particle masses successively larger by a factor of $8$ for each layer, 
which then connects with
the outer root grid that has a dark matter particle mass $8^3$ times that in the refined region.
The initial density fluctuations 
are included up to the Nyquist frequency in the refined region.
The surrounding volume outside the refined region 
is also followed hydrodynamically, which is important in order to properly capture
matter and energy exchanges at the boundaries of the refined region.
Because we still can not run a very large volume simulation with adequate resolution and physics,
we choose these two runs of moderate volumes to represent two opposite environments that possibly bracket the universal average.

We choose a varying mesh refinement criterion scheme such that the resolution is always better than $460$/h proper parsecs 
within the refined region, corresponding to a maximum mesh refinement level of $9$ above $z=3$, 
of $10$ at $z=1-3$ and $11$ at $z=0-1$.
The simulations include a metagalactic UV background
\citep[][]{2012Haardt},  
and a model for shielding of UV radiation \citep[][]{2005Cen}.
The simulations also include metallicity-dependent radiative cooling and heating \citep[][]{1995Cen}. 
The Enzo version used includes metallicity-dependent radiative cooling extended down to $10$K,
molecular formation on dust grains,
photoelectric heating and other features that are different from or not in the public version of Enzo code.
We clarify that our group has included metal cooling and metal heating (due to photoionization of metals) 
in all our studies since \citet[][]{1995Cen} for the avoidance of doubt \citep[e.g.,][]{2009Wiersma, 2011TepperGarcia}.
Star particles are created in cells that satisfy a set of criteria for 
star formation proposed by \citet[][]{1992CenOstriker}.
Each star particle is tagged with its initial mass, creation time, and metallicity; 
star particles typically have masses of $\sim$$10^{5-6}\msun$.

Supernova feedback from star formation is modeled following \citet[][]{2005Cen}.
Feedback energy and ejected metal-enriched mass are distributed into 
27 local gas cells centered at the star particle in question, 
weighted by the specific volume of each cell (i.e., weighting is equal to the inverse of density), 
which is to mimic the physical process of supernova
blastwave propagation that tends to channel energy, momentum and mass into the least dense regions
(with the least resistance and cooling).
We allow the whole feedback processes to be hydrodynamically coupled to surroundings
and subject to relevant physical processes, such as cooling and heating, as in nature.
The extremely inhomogeneous metal enrichment process
demands that both metals and energy (and momentum) are correctly modeled so that they
are transported into right directions in a physically sound (albeit still approximate 
at the current resolution) way, at least in a statistical sense.
In our simulations metals are followed hydrodynamically by solving 
the metal density continuity equation with sources (from star formation feedback) and sinks (due to subsequent star formation).
Thus, metal mixing and diffusion through advection, turbulence and other hydrodynamic processes
are properly treated in our simulations.

The primary advantages of this supernova energy based feedback mechanism are three-fold.
First, nature does drive winds in this way and energy input is realistic.
Second, it has only one free parameter $e_{SN}$, namely, the fraction of the rest mass energy of stars formed
that is deposited as thermal energy on the cell scale at the location of supernovae.
Third, the processes are treated physically, obeying their respective conservation laws (where they apply),
allowing transport of metals, mass, energy and momentum to be treated self-consistently 
and taking into account relevant heating/cooling processes at all times.
We use $e_{SN}=1\times 10^{-5}$ in these simulations.
The total amount of explosion kinetic energy from Type II supernovae
with a Chabrier IMF translates to $e_{SN}=6.6\times 10^{-6}$.
Observations of local starburst galaxies indicate
that nearly all of the star formation produced kinetic energy (due to Type II supernovae)
is used to power galactic superwinds \citep[e.g.,][]{2001Heckman}.
Given the uncertainties on the evolution of IMF with redshift (i.e., possibly more top heavy at higher redshift)
and the fact that newly discovered prompt Type I supernovae contribute a comparable
amount of energy compared to Type II supernovae, it seems that our adopted value for
$e_{SN}$ is consistent with observations and physically realistic.
The validity of this thermal energy-based feedback approach comes empirically.
In \citet[][]{2012Cen} the metal distribution in and around galaxies over a wide range of redshift
($z=0-5$) is shown to be in excellent agreement with respect to the properties of observed damped $\lya$ systems
\citep[][]{2012Rafelski},
whereas in \citet[][]{2012bCen} we further show that 
the properties of OVI absorption lines at low redshift, including their abundance, Doppler-column density distribution,
temperature range, metallicity and coincidence between OVII and OVI lines,
are all in good agreement with observations \citep[][]{2008Danforth,2008Tripp, 2009Yao}.
This is non-trivial by any means, because they require that the transport of metals and energy from galaxies to
star formation sites to megaparsec scale be correctly modeled as a function of distance over the entire cosmic timeline,
at least in a statistical sense.

\subsection{Analysis Method}

We identify galaxies at each redshift in the simulations using the HOP algorithm
\citep[][]{1998Eisenstein} operating on the stellar particles, which is tested to be robust
and insensitive to specific choices of concerned parameters within reasonable ranges.
Satellites within a galaxy down to mass of $\sim 10^9\msun$ are clearly identified separately in most cases.
The luminosity of each stellar particle in each of the Sloan Digital Sky Survey (SDSS) five bands
is computed using the GISSEL stellar synthesis code \citep[][]{Bruzual03},
by supplying the formation time, metallicity and stellar mass.
Collecting luminosity and other quantities of member stellar particles, gas cells and dark matter particles yields
the following physical parameters for each galaxy:
position, velocity, total mass, stellar mass, gas mass,
mean formation time,
mean stellar metallicity, mean gas metallicity, SFR,
luminosities in five SDSS bands (and various colors) and others.
We show, among others, that the simulated luminosity functions of galaxies at $z=0$ are reasonably 
matched to observations \citep{2011Cen}.

In the analysis presented here we choose randomly ten galaxies from our simulation that have properties 
that are similar to observed galaxies in the COS-HALO program \citep{2013Werk, 2014Werk}
with respect to the star formation rate (SFR) and stellar mass.
Some relevant properties of the ten simulated galaxies are tabulated in Table 1. 
A central galaxy is defined to be one that is not within the virial radius of a larger (halo-mass-wise) galaxy.

\begin{table}[ht!]
\caption{Properties of 10 simulated galaxies used in this study.}
\centering
\begin{tabular}{cccccc}
\hline\hline 
Stellar M [$10^{10}\msun$] & Halo M [$10^{11}\msun$] & SFR [$\msun/{\rm yr}$]&<T>[$10^{4}$K]&<Z>({\rm gas})[$\zsun$]& central galaxy\\ 
1.37 & 1.37 & 1.0 & 27.95 & 0.23 & yes\\
1.56 & 1.17 & 1.5 & 13.77 & 0.31 & yes\\
2.25 & 10.4 & 1.6 & 36.76 & 0.20 & no \\
3.00 & 4.36 & 1.7 & 16.89 & 0.27 & no \\
2.91 & 2.05 & 1.0 & 10.10 & 0.23 & yes \\
3.03 & 2.45 & 1.6 & 14.74 & 0.30 & yes \\
3.63 & 3.89 & 2.2 & 93.48& 0.27 & yes \\
3.24 & 4.92 & 1.8 & 25.05& 0.16 & yes \\
3.68 & 6.22 & 2.6 & 15.44& 0.18 & yes \\
3.97 & 3.19 & 1.6 & 20.79& 0.13 & yes \\\hline
\end{tabular}
\label{table:properties}
\end{table}

To assist performing post-simulation analysis of the galaxies,
we construct lookup tables of the abundances of various ions 
of elements nitrogen, oxygen and silicon 
as a function of logarithm of temperature ($\log T$) and logarithm of ionization parameter ($\log U$),
for solar metallicity, using the photoionization code CLOUDY \citep[v13.03;][]{2013CLOUDY}. 

For each selected simulated galaxy,
we construct a cube with size of 320 kpc centered on the galaxy with resolution of 625 pc.
We make the simplifying but reasonable assumption that
the relevant absorbers are optically thin.
Our calculations are performed for two cases of ionizing radiation field that the CGM
in each galaxy is assumed to be exposed to.
In the first case, we only use 
the HM background UV radiation field at $z=0$.  
In the second case, 
we compute the ionizing UV radiation due to local, shocked heated gas within each concerned galaxy and
use the sum of that and the HM background.

The local UV ionizing radiation is computed as follows.
We compute the emissivity ($e_{\nu}$) [${\rm erg s^{-1} cm^3 Hz^{-1} sr^{-1}}$]
for each cell given its temperature and metallicity at 
the relevant energies, 
$E=47.3$eV for SiIV,  $E=97.88$ eV for NV, $E=47.4$ eV for NIII, $E=29.6$ eV for NII and $E=138.1$ eV for OVI ion.
This is done by integrating the diffuse spectrum from CLOUDY between $97.88\,eV, 1.2\times97.88\,eV$ for NV, as an example
(similarly for other ions). 
The diffuse emission includes all gas processes, the free-free emission, radiative free-bound recombination, 
two-photon emission, and electron scattering, among others, for all elements in the calculation.
For each cell the total luminosity density is computed as 
$n^2\times \Delta V\times L_{\nu}$ where $n$ is the density and $\Delta V$ its volume. 
The sum of local ionizing UV radiation luminosity density at a relevant wavelength is $L_\lambda$.
To approximately account for the spatial distribution of local UV radiation sources without
the expense of detailed radiative transfer,
we compute the half luminosity radius (${\rm R_e}$) of a galaxy, within which half of the local
radiation luminosity in that galaxy originates. 
Then, we assign the local flux to each cell with distance $r$ from the center of the galaxy, approximately, as 
\begin{equation}
\label{eq:F}
{\rm F_{\lambda,r}=\frac {L_{\lambda}}{4 \pi r^2} [1+2e^{-r/(2R_e)}]}.
\end{equation}
\noindent 
The new ionization radiation ``background" at each cell with the inclusion of the local emission is computed as 
$F_{new}= F_{HM,\lambda} + F_{r,\lambda}$,
where $F_{HM,\lambda}$ is the flux of the HM background radiation at the relevant wavelength $\lambda$. 
Since the local radiation is mostly dominated by dense hot gaseous regions that tend to be spatially centrally 
concentrated, our neglect of its possible attenuation likely makes the second case of radiation field
(HM+local) an upper limit.
Thus, the two choices of radiation field likely bracket all possible cases.

Each cell has a size of 625 pc within a cube of 320 kpc centered on each selected simulated galaxy, 
we convert the density $n_H$ and the radiation $F$ at the cell 
to the ionization parameter $U=F/c n_{H}$ at the radiation energy in question, where $c$ is speed of light.
Using $U$ and temperature of the cell,
we find the abundances of various ions for each cell 
using the pre-computed CLOUDY lookup tables,
which is then multiplied by the metallicity of the cell in solar units.
We use the updated solar abundances of these elements from \citet{2009Asplund}, 
in the notation of $\rm \log \epsilon_{x}=\log (N_x/N_H)+12$ listed in Table 2.
Table 2 lists the UV lines analyzed in this paper,
where each doublet is listed using two rows.
The information for each line is listed, including wavelength (column 2),
oscillator strength (column 3),
lower and upper energy levels of the transition (columns 4,5)
and abundance of the element (column 6).
In column 7, we list the lower column density threshold in constructing
covering fractions of the lines (see Figure \ref{fig:cover}).
Each of the lower column density thresholds is chosen to be 
the minimum of the upper limits for each respective ion. 
For computing the frequency of the line ratios, essentially some line to the OVI line in all cases, 
we choose the cut for the OVI column density at $\rm \log N_{OVI}>14$, 
which approximately corresponds to the lowest column density of detected OVI absorbers.
Note that all of the lines studied here are resonant lines.

\begin{table}[H]
\caption{Properties of UV lines analyzed in this study.}
\centering
\begin{tabular}{c c c c c c c}
\hline\hline
Ion & wavelength[\AA]&oscillator strength& lower level&upper level&$\log\epsilon$ & ${\rm \log N_{cut}}$ \\
\hline
NV & 1238.8 & 1.56e-1 & $^2S_{1/2}$ & $^2P_{1/2}$&7.83 & 13.42\\
NV & 1242.8 & 7.8e-2 & $^2S_{1/2}$ & $^2P_{3/2}$&7.83 & 13.42 \\
NIII & 989.7 & 1.23e-1 & $^2P_{1/2}$ & $^2D_{3/2}$&7.83 & 13.50\\
NII & 1083.9 & 1.11e-1 & $^3P_{0}$ & $^3D_{1}$&7.83 & 13.46\\
OVI & 1031.9 & 1.33e-1 & $^2S_{1/2}$ & $^2P_{3/2}$&8.69 & 13.27 \\
OVI & 1037.6 & 6.6e-2 & $^2S_{1/2}$ & $^2P_{1/2}$&8.69 & 13.27 \\
SiIV & 1393.7 & 5.13e-1 & $^2S_{1/2}$ & $^2P_{3/2}$&7.51 & 12.38 \\
SiIV & 1402.7 & 2.55e-1 & $^2S_{1/2}$ & $^2P_{1/2}$&7.51 & 12.38 \\\hline\end{tabular}\label{table:ions}\end{table}

Unlike oxygen and silicon, nitrogen stems from both primary and secondary producers
and consequently nitrogen abundance is theoretically expected to be a function of overall metallicity,
for which oxygen abundance is a good proxy.
This theoretical expectation is confirmed by observations.
We use the
fitting formula of \citet{2006Molla}, which is normalized at solar value,  
to express the ${\rm N/O}$ ratio as a function of ${\rm O/H}$ ratio:
\begin{equation}
\label{eq:NO}
{\rm \log(N/O) = -1149.31 + 1115.23 x - 438.87x^2 + 90.05x^3 -10.20x^4 + 0.61x^5 - 0.015x^6},
\end{equation}
where $\rm x = 12 + \log (O/H)$.
In subsequent analysis, where nitrogen is concerned,
we perform the analysis twice, one assuming ${\rm N/O}$ to be independent of ${\rm O/H}$ and 
another using Eq (\ref{eq:NO}). 
To give the magnitude of the effect, we note that
at $(0.03, 0.1, 0.3)\zsun$ for oxygen abundance,
${\rm N/O}$ value is $(0.27,0.28,0.4)$ in solar units.

\section{Results}

\begin{figure}[h!]
\centering
\vskip -0.3cm
\resizebox{6.0in}{!}{\includegraphics[angle=0]{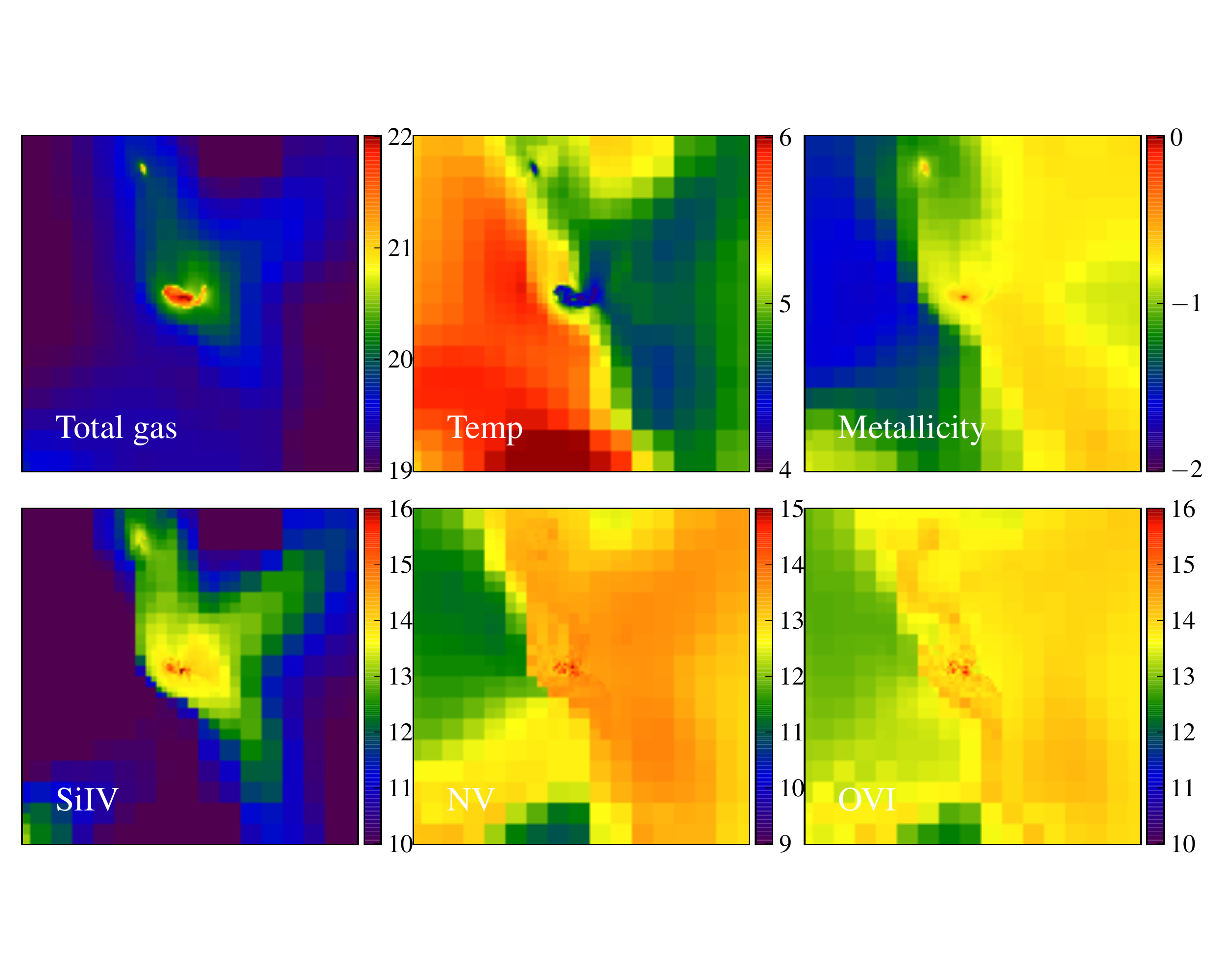}}
\vskip -1.5cm
\caption{
shows projection plots along one of the axes of the 320 kpc cube for one of the galaxies listed in Table 1.
From top-left in clockwise direction are logarithm of total hydrogen column density (top-left), 
logarithm of the density-weighted gas temperature (top-middle),
logarithm of the density-weighted gas metallicity in solar units (top-right),
logarithm of OVI column density (bottom-right),
logarithm of NV column density (bottom-middle)
and
logarithm of SiIV column density (bottom-left).}
\label{fig:pic1}
\end{figure}

\begin{figure}[h!]
\centering
\vskip -1.5cm
\resizebox{6.0in}{!}{\includegraphics[angle=0]{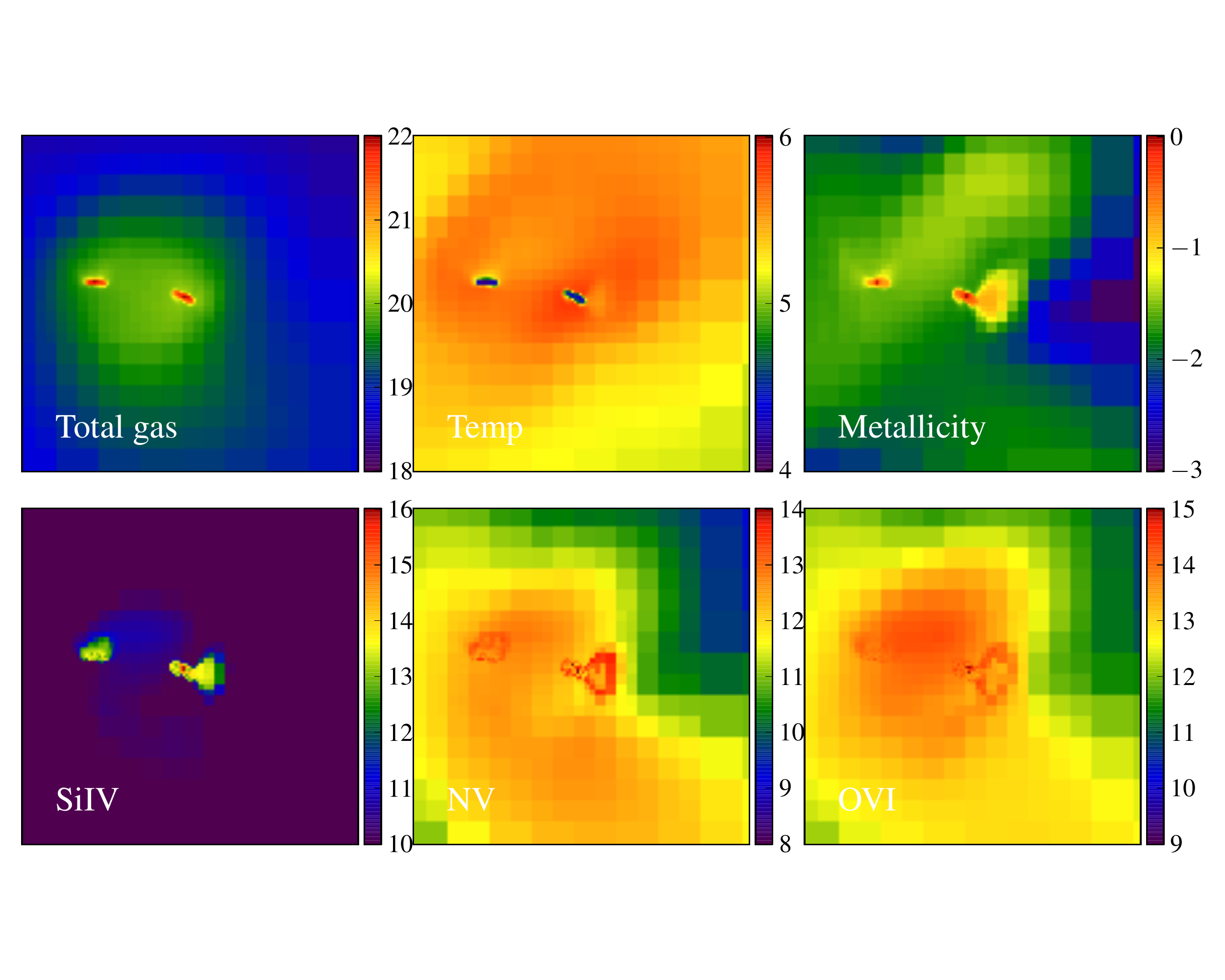}}
\vskip -1.5cm
\caption{
shows the same as in Figure \ref{fig:pic1} but for another galaxy.
}
\label{fig:pic2}
\end{figure}

\begin{figure}[!h]
\centering
\vskip -0.4cm
\resizebox{6.0in}{!}{\includegraphics[angle=0]{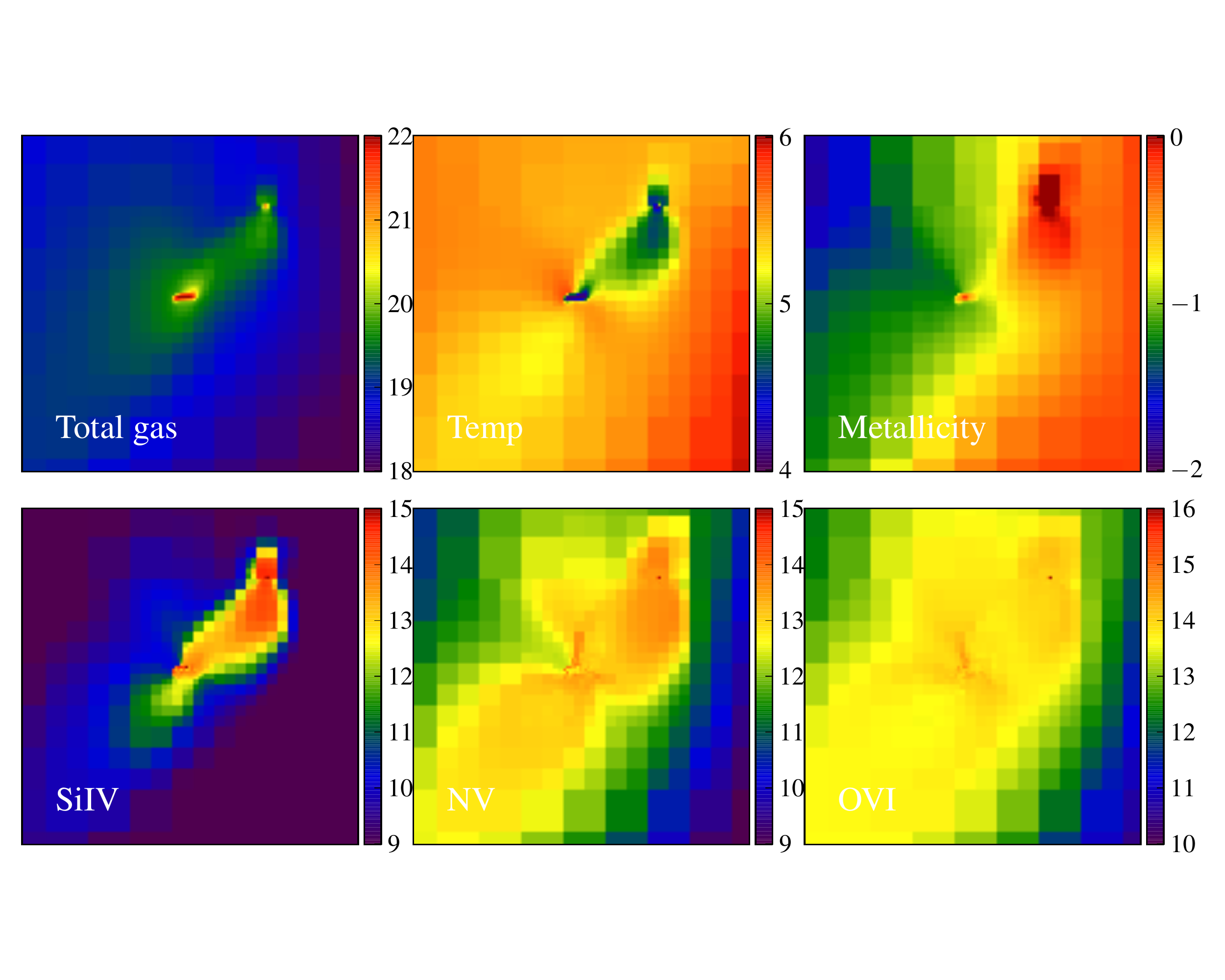}}
\vskip -1.5cm
\caption{
shows the same as in Figure \ref{fig:pic1} for yet another galaxy.
}
\label{fig:pic3}
\end{figure}

Before presenting quantitative results,
we show visually some basic quantities for a few galaxies
in Figures\ref{fig:pic1}, \ref{fig:pic2}, \ref{fig:pic3}.
Some features are easily visible just from these three random examples.
First, large variations from galaxy to galaxy are evident,
for each of the displayed variables.
Physically, this stems from density and thermodynamic structures of each galaxy being 
subject to its unique exterior and interior forces, including 
halo mass, gas inflow and associated dynamic and thermodynamic effects,
feedback from star formation and related dynamic and thermodynamic effects,
As an illustrative example, 
in Figure\ref{fig:pic1}, 
we see the temperature at the lower-left triangle mostly in the range of $10^{5.5}-10^6$K,
compared to the temperature at the upper-right triangle mostly in the range of $10^{4.5}-10^5$K.
We do not investigate here further into the dynamic causes of such temperature patterns
with possible physical processes including merger shocks and stellar feedback (i.e., supernova) shocks.
Second, the temperature distribution of the CGM is far from uniform.
Indeed, the CGM is of multi-phase in nature,
typically spanning the range of $10^4-10^6K$ within the $\sim 150$ kpc radial range for the galaxies examined.
This property is of critical importance to the line ratios that we obtain in the simulations.
Third, the metallicity distribution 
in the CGM is highly inhomogeneous,
typically spanning $10^{-2}-10^0\zsun$.
Fourth, although the number of galaxies looked at is small, 
we find that star-forming galaxies, as those selected in this investigation,
tend to be involved in significant mergers.
This in turn suggests that significant mergers may be a necessary
ingredient in driving significant star formation activities in galaxies at low redshift.

\begin{figure}[!h]
\centering
\hskip -1.0cm
\resizebox{3.3in}{!}{\includegraphics[angle=0]{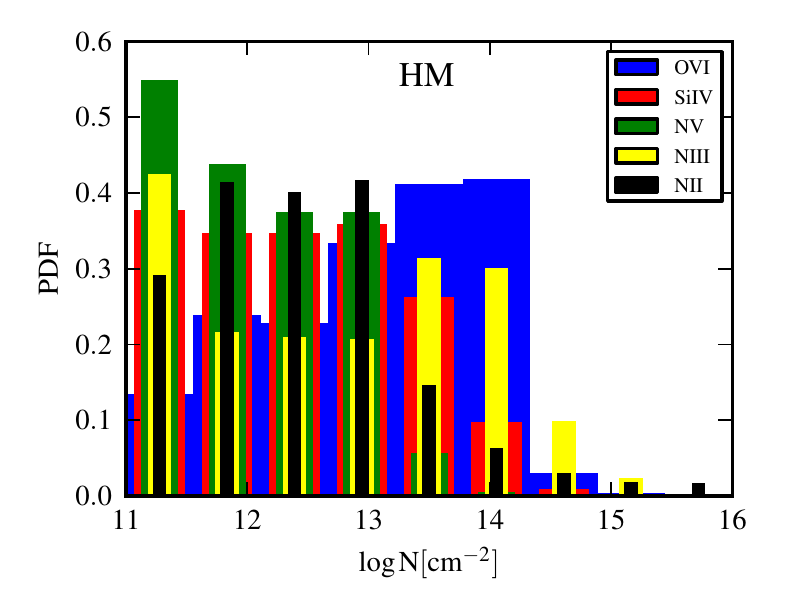}}
\hskip -0.3cm
\resizebox{3.3in}{!}{\includegraphics[angle=0]{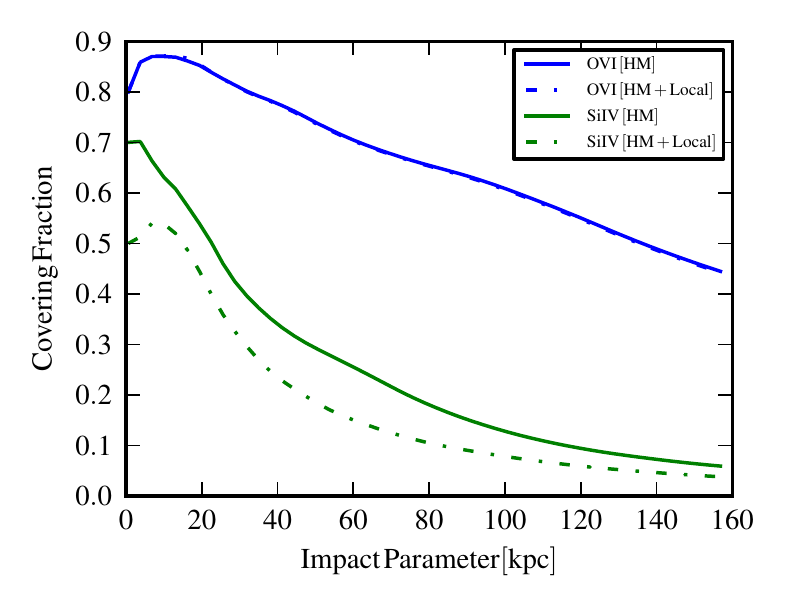}}
\vskip -0.0cm
\hskip -1.0cm
\resizebox{3.3in}{!}{\includegraphics[angle=0]{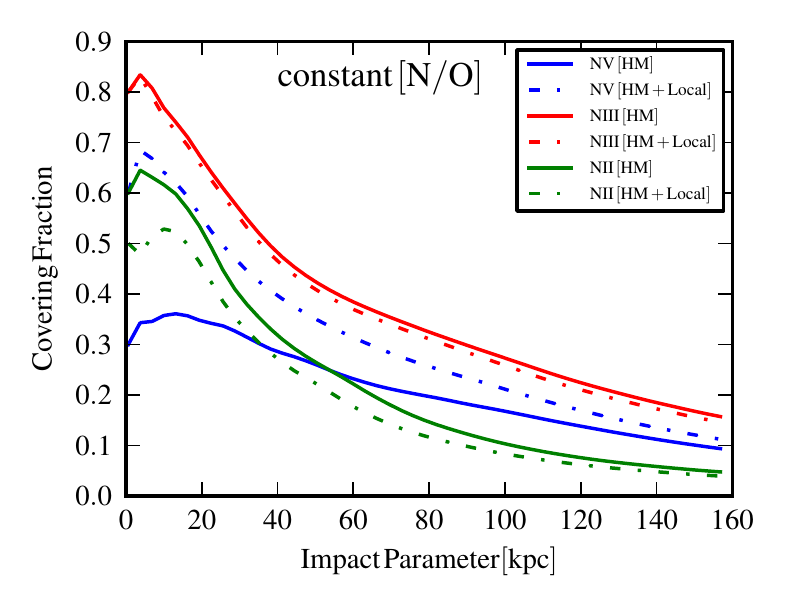}}
\hskip -0.3cm
\resizebox{3.3in}{!}{\includegraphics[angle=0]{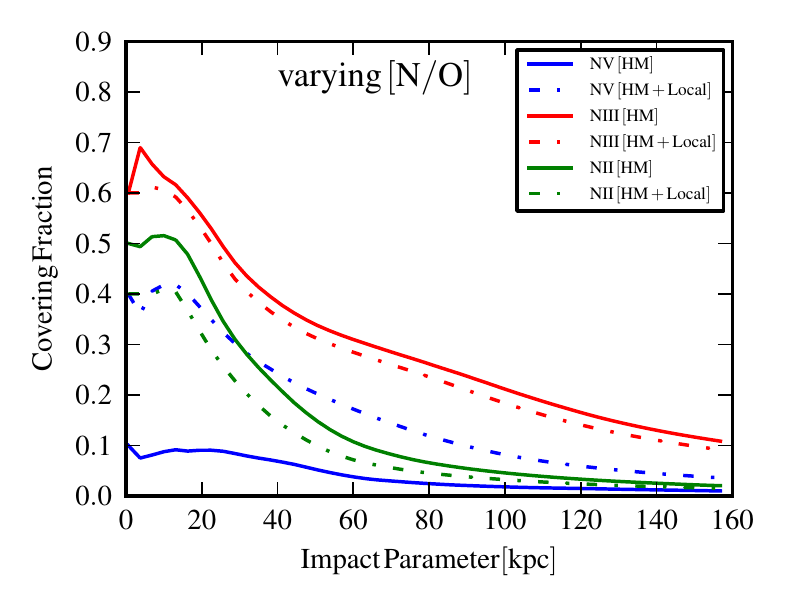}}
\vskip -0.0cm
\caption{
{\color{burntorange}\bf Top-left panel:}
shows the column density distributions for all five lines in the case with HM,
and for the three nitrogen lines with varying $\rm N/O$ (Eq~\ref{eq:NO}).
{\color{burntorange}\bf Top-right panel:}
 shows the covering fraction as a function of the galacto-centric impact parameter
for OVI absorbers with column density 
above $10^{13.27}$cm$^{-2}$ with HM (solid blue curve) and with HM+local (dot-dashed blue curve),
for SiIV absorbers with column density 
above $10^{12.38}$cm$^{-2}$ with HM (solid green curve) and with HM+local (dot-dashed green curve).
{\color{burntorange}\bf
Bottom-left panel:}
shows the same as 
in the top panel but for the three nitrogen absorption lines 
with column density above $10^{13.43}$cm$^{-2}$ for NV (blue curves), 
$10^{13.50}$cm$^{-2}$ for NIII (red curves)
and $10^{13.46}$cm$^{-2}$ for NII (green curves), 
under the assumption that $\rm N/O$ ratio is independent of metallicity.
The solid curves correspond to the case with only HM radiation background,
whereas the dot-dashed curves are for the case with HM+local radiation.
{\color{burntorange}\bf
Bottom-right panel:}
shows the same as for the bottom-left panel,
except that we use Eq (\ref{eq:NO}) for nitrogen abundance as a function of oxygen abundance.
}
\label{fig:cover}
\end{figure}

We now turn to quantitative results.
The top-left panel of Figure \ref{fig:cover} shows the column density distributions
of the five lines. 
Due to our projection method, the number of weaker lines are underestimated due to blending.
The turndown of the number of the OVI lines below column density $10^{13}$cm$^{-2}$ is probably due to that.
This is unlikely to significantly affect our results below, since our coincidence analysis 
is focused on OVI absorbers with column density higher than $10^{14}$cm$^{-2}$. 
The covering fractions shown in Figure \ref{fig:cover} may be somewhat overestimated,
since the column density cutoff for OVI is $10^{13.27}$cm$^{-2}$;
a comparison between the cutoff column densities listed in Table \ref{table:ions}
and the behavior of the column density histograms shown in the top-left panel of
Figure \ref{fig:cover} for the other four lines (SiIV, NV, NIII and NII) suggests
that the covering fractions for these four lines are unlikely affected significantly due to blending.

\begin{figure}[!h]
\centering
\vskip -1.0cm
\hskip -1.0cm
\resizebox{3.3in}{!}{\includegraphics[angle=0]{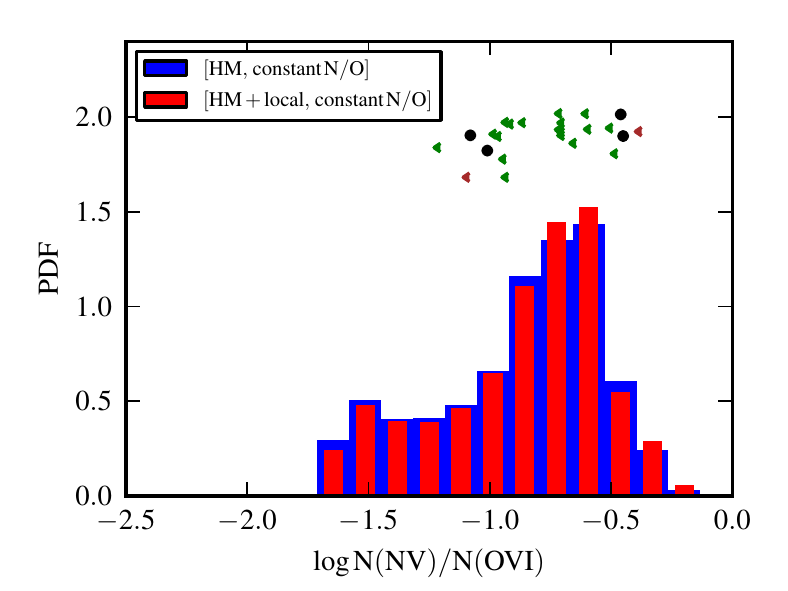}}
\hskip -0.0cm
\resizebox{3.3in}{!}{\includegraphics[angle=0]{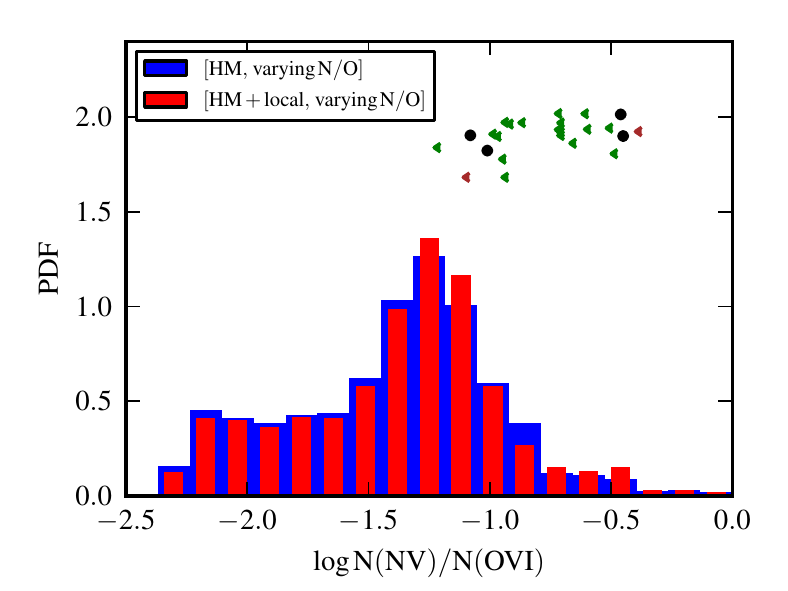}}
\vskip -0.0cm
\hskip -1.0cm
\resizebox{3.3in}{!}{\includegraphics[angle=0]{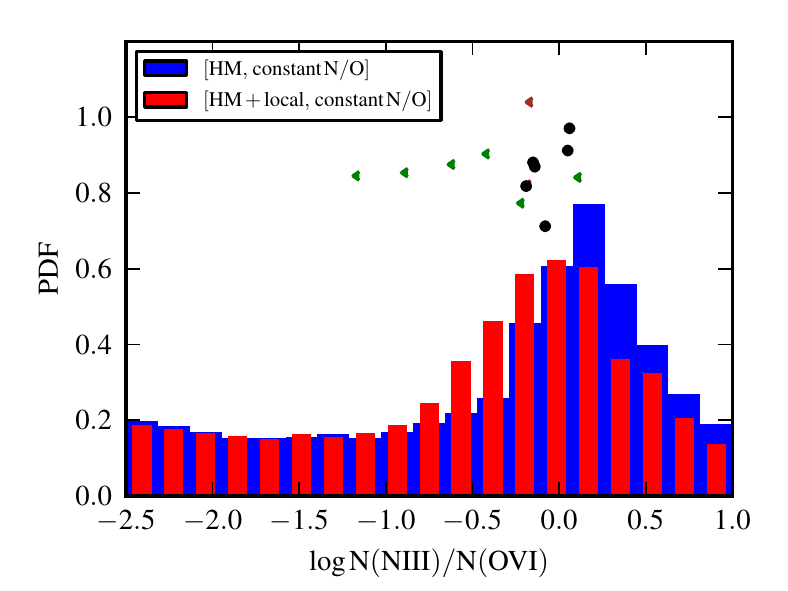}}
\hskip -0.0cm
\resizebox{3.3in}{!}{\includegraphics[angle=0]{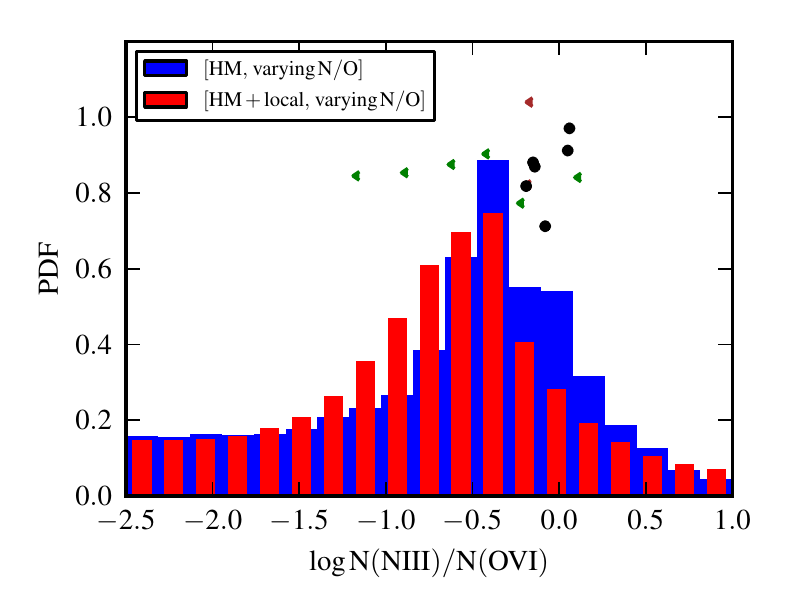}}
\vskip -0.0cm
\hskip -1.0cm
\resizebox{3.3in}{!}{\includegraphics[angle=0]{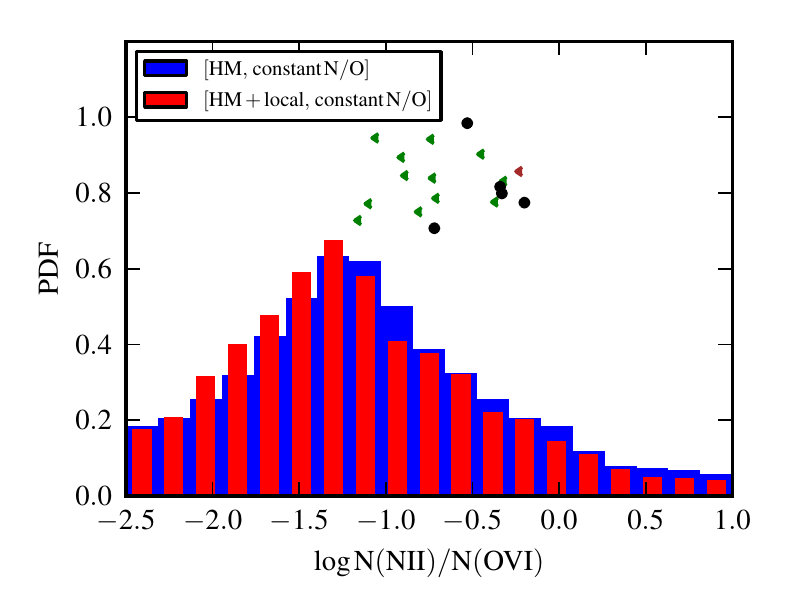}}
\hskip -0.0cm
\resizebox{3.3in}{!}{\includegraphics[angle=0]{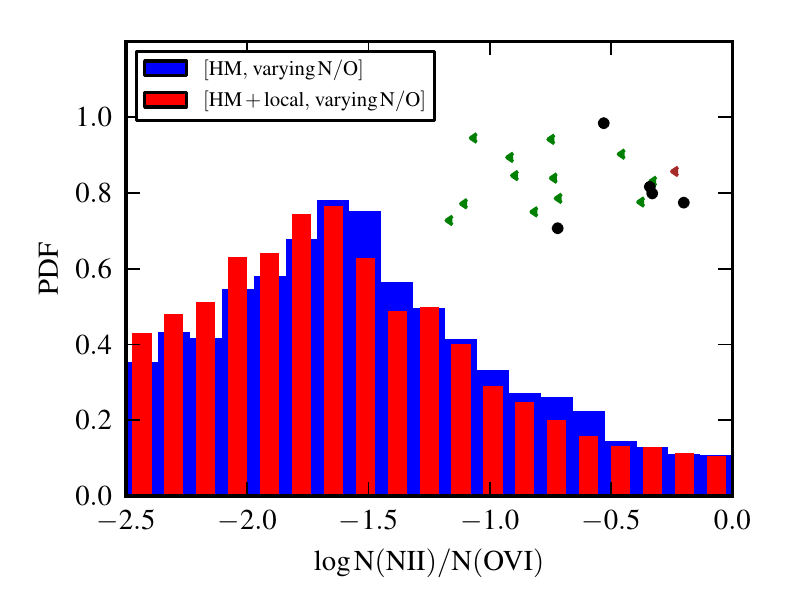}}
\vskip -0.5cm
\caption{
{\color{burntorange}\bf Top row:}
shows the probability distribution function (PDF) of the ratio of ${\rm N(NV)/N(OVI)}$
for all OVI absorbers with $\rm N(OVI)>10^{14}$cm$^{-2}$ with constant N/O ratio (left)
and varying N/O as a function of O/H (Eq \ref{eq:NO}).
{\color{burntorange}\bf Middle row:} the same for NIII/OVI.
{\color{burntorange}\bf Bottom row:} the same for NII/OVI.
(Blue, red) histograms are for (HM, HM+local) radiation field. 
The observational data are shown for three separate types:
black dots are those with both lines detected,
left green arrows are those where the numerator line is an upper limit and the denominator line a detection,
and
brown left arrows are those where the numerator line is an upper limit and the denominator line a lower limit.
The y coordinates of the points are arbitrary. 
}
\label{fig:Nhist}
\end{figure}

The remaining three panels of Figure \ref{fig:cover} show covering fraction of OVI and SiIV (top-right panel),
NV, NIII, NII lines with constant N/O ratio (bottom-left panel) 
and with N/O as a function of O/H (Eq~\ref{eq:NO}, bottom-right panel).
Several interesting properties may be noted.
First, there is a significant drop of covering fraction,
by a factor of $2-10$, from the central regions (a few kpc) to $\sim 150$ kpc.
This is likely due primarily to a combination of the general trend of decreasing gas density and decreasing
metallicity of the CGM with increasing galacto-centric radius.
In spite of this covering fraction decrease with radius,
most of the absorbers are located at large impact parameters,
since the area increases with radius at a higher rate, for example, by a factor of 64 from
$20$ kpc to $160$ kpc.
Second, the OVI covering fraction (top panel)
is large and largest among the examined lines, ranging from 
$80-90$\% at $\le 10$ kpc to $\sim 50$\% at $\le 150$ kpc,
given the chosen column density thresholds listed in Table \ref{table:ions}.
This is in good agreement with observations 
\citep[e.g.,][]{2009Chen,2011bProchaska}.
Third, it is particularly noteworthy that there is essentially no difference between 
HM and HM+local cases for the OVI covering fraction.
This indicates that photoionization plays a negligible role in the abundance of OVI.
In other words, OVI is produced by collisional processes, powered by feedback and gravitational shocks,
which will be verified subsequently
[see \citet{2013Cen} for a detailed discussion on the varying contributions of stellar feedback versus gravitational shocks
in different types of galaxies].
Fourth, a stronger radiation field tend to increase the abundance of NII and NIII but the opposite is true for NV.
But the difference between HM and HM+local cases for both NII and NIII are fairly minor, indicating
that collisional processes are the primary powering source for producing NII and NIII.
This is not the case for SiIV and NV, where
the differences between HM and HM+local cases are substantial 
and the differences are larger toward small impact parameters.
This suggests that a higher HM+local radiation is able to produce NV for high density
gas in the inner regions of star-forming galaxies, while the outer regions are mainly 
dominated by collisional processes, consistent with a trend of increasing temperature 
with increasing galacto-centric radius found in \citet{2013Cen}.
Finally, comparing the bottom-left and bottom-right panels,
we see significant differences between constant N/O case (left) 
and varying N/O case (Eq \ref{eq:NO}, right).
A closer look reveals that the difference increases with increasing impact parameter,
reflecting the trend of decreasing gas metallicity ($\rm O/H$) with increasing impact parameter.
Also revealed is that the decreases in covering fraction
from constant N/O to varying N/O case for different lines differ significantly,
reflecting the complex multi-phase medium with inhomogeneous, temperature-and-density-dependent metallicity distribution;
while the decrease for NIII is relatively small (a factor of less than two for all impact parameters),
the decreases for NII and NV are quite large, a factor of larger than two at $<150$ kpc.
As we will quantify using KS tests subsequently, the significant reduction in nitrogen abundance
with the varying N/O case results in noticeably better KS test p-values with respect to 
NII/OVI, NIII/OVI, NV/OVII column density ratios.

\begin{figure}[!h]
\centering
\vskip -0.0cm
\resizebox{3.3in}{!}{\includegraphics[angle=0]{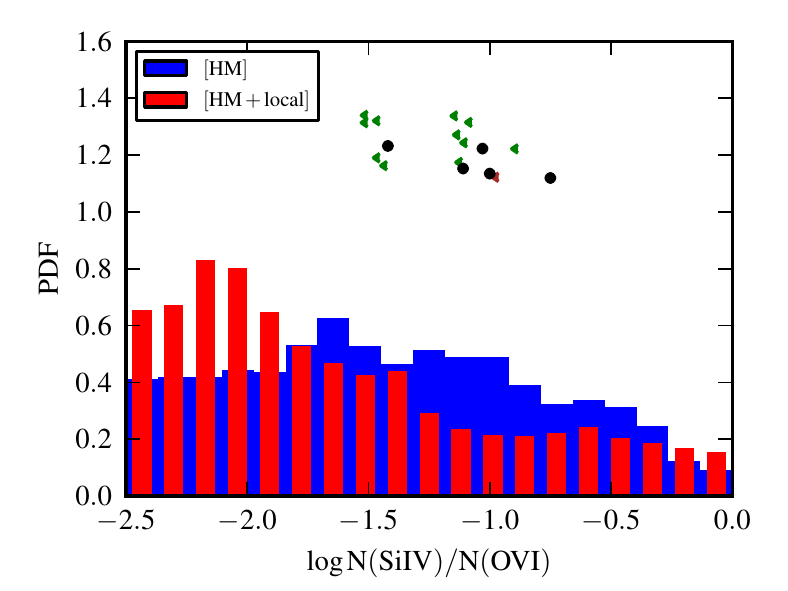}}
\vskip -0.0cm
\caption{
shows the PDF of the ratio of $\rm N(SiIV)/N(OVI)$
for all OVI absorbers with $\rm N(OVI)>10^{14}$cm$^{-2}$.
The points are observational data divided into three separate types:
the black dots are those with both lines in the ratio detected,
the left green arrows are those where SiIV line is an upper limit and OVI line is a detection,
and
the brown left arrows are those where SiIV line is an upper limit and OVI line is a lower limit.
The blue histograms are for HM radiation field, whereas the red histograms are for HM+local radiation field.
The y coordinates of the points are arbitrary.}
\label{fig:SiIVhist}
\end{figure}

We now make direct comparisons to observations with respect to the ratio of column densities for four absorption line pairs.
Figure \ref{fig:Nhist} 
shows the probability distribution functions 
of ${\rm N(NV)/N(OVI)}$ (top row),
${\rm N(NIII)/N(OVI)}$ (middle row)
and 
${\rm N(NII)/N(OVI)}$ (bottom row),
with each row further separated for constant N/O (left) and varying N/O cases (right).
Due to the fact that the vast majority of NII, NIII and NV absorbers 
have metallicities in the range of $\rm [O/H]=-1$ to $-0.5$ (see Figure \ref{fig:contour_ZN} below),
the horizontal shifts of the peaks of the PDFs for the ratios of all three nitrogen lines to OVI
are substantial, of order 0.5 dex.
The shifts have noticeable effects on the KS tests between simulations and observations below
given in Table \ref{table:KS}.
An examination by eye between the simulation results and observational data 
comes with the visual impression that all cases agree reasonably well with observations, 
which will be verified quantitatively.
Figure \ref{fig:SiIVhist} 
shows the probability distribution functions 
of $\rm N(SiIV)/N(OVI)$ column density ratios.
Once again, visual examination suggests agreement with simulations and observations.

\begin{table}[!h]
\caption{Two-sample KS test p-values for column density ratio distributions of four absorption line pairs, 
including cases with constant and varying N/O ratios and HM versus HM+local radiation field}
\centering
\begin{tabular}{c c c}
\hline\hline
Line ratio & HM & HM+Local\\
\hline
NV/OVI[constant N/O]& 0.33 & 0.37  \\
NII/OVI[constant N/O]&0.93&0.99 \\
NIII/OVI[constant N/O]&0.28&0.49\\
NV/OVI[varying N/O]& 0.99 & 0.99  \\
NII/OVI[varying N/O]&0.99&0.99 \\
NIII/OVI[varying N/O]&0.91&0.96\\
SiIV/OVI& 0.9 & 0.98\\ 
\hline
\end{tabular}
\label{table:KS}
\end{table}

To gain a more quantitative statistical test between simulations and observations,
we perform two-sample KS tests between simulated and observed column density ratios for four pairs of lines,
NV/OVI, NII/NV, NIII/OVI and SiIV/OVI.
Since most of the observational data are upper and lower limits, instead of actual detections,
our analysis is performed as follows.
For the case where the numerator line is an upper limit and the denominator line is a detection,
we allow the ratio to be drawn from the simulation distribution with value upper-bounded by the upper limit.
The same is done for the case where the numerator line is an upper limit and the denominator line is a lower limit.
Then, in conjunction with detections, where both lines are detected,
we perform a two-sample KS test for each of the four line pairs, NV/OVI, NII/NV, NIII/OVI and SiIV/OVI,
between simulations and observations.
Needless to say, our presently adopted procedure to treat the upper and lower limits cases
favors agreement with observations and simulations.
Nevertheless, the procedure is consistent with the current data.
The results are tabulated in Table \ref{table:KS}.
Clearly, no major disagreements can be claimed as to reject the simulation results
in all four cases, (constant N/O, varying N/O) times (HM, HM+local).
However, there are hints, at face value, that the constant N/O cases are less preferred
than the varying N/O cases.
Nonetheless, it is premature to make any firm statistical conclusion on that at this juncture.
Thus, the only robust conclusion we can reach at this time is that
our simulation predictions are fully consistent with extant observational data
with respect to the four line ratios, NV/OVI, NII/NV, NIII/OVI and SiIV/OVI.

What would exponentially increase the statistical power of testing the models 
is to turn these current upper limits into real detections.
We have performed the following exercise to demonstrate this point.
Let us assume that all current upper limits of the column density ratios become
detections and the detection values are lower than current upper limits uniformly by a factor of $\Delta$ dex.
We find that, 
if $\Delta=(0.18,0.16)$,
the KS p-values for the NV/OVI line ratio become
$(0.05,0.05)$ for the (HM, HM+local) cases with constant N/O;
with $\Delta=(0.21,0.20)$, the KS p-values for the NV/OVI line ratio become
$(0.01,0.01)$ for the (HM, HM+local) cases with constant N/O.
For the varying N/O cases, the situation is non-monotonic in the following sense:
the KS p-values for the NV/OVI line ratio are $(0,0)$ for the (HM, HM+local) with $\Delta=(0,0)$, 
increasing to a maximum of $(0.5,0.8)$ with $\Delta=(0.70,0.63)$, 
then downturning to $(0.01,0.01)$ with $\Delta=(0.83,0.78)$. 
Obviously, a uniform shift is an oversimplification.
Nevertheless, this shows clearly an urgent need to increase observational sensitivity in order to place 
significantly stronger constraints on models than currently possible.
When all line pairs are deployed, the statistical power will be still, likely much, greater.

\begin{figure}[!h]
\centering
\vskip -0.0cm
\hskip -1.0cm
\resizebox{3.4in}{!}{\includegraphics[angle=0]{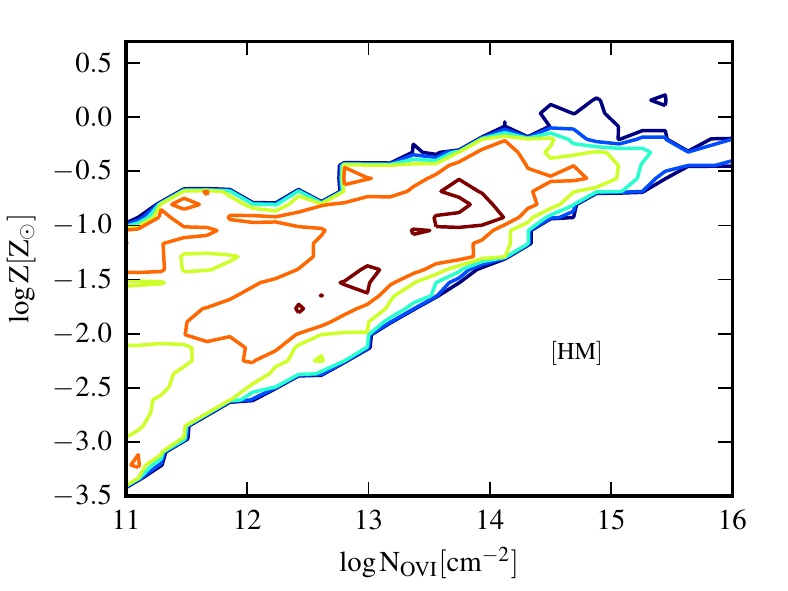}}
\hskip -0.0cm
\resizebox{3.4in}{!}{\includegraphics[angle=0]{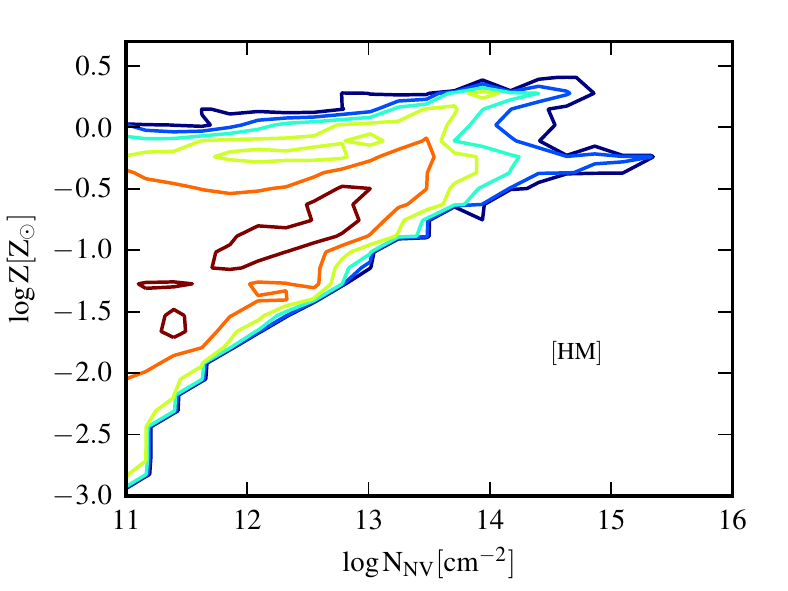}}
\vskip -0.0cm
\hskip -1.0cm
\resizebox{3.4in}{!}{\includegraphics[angle=0]{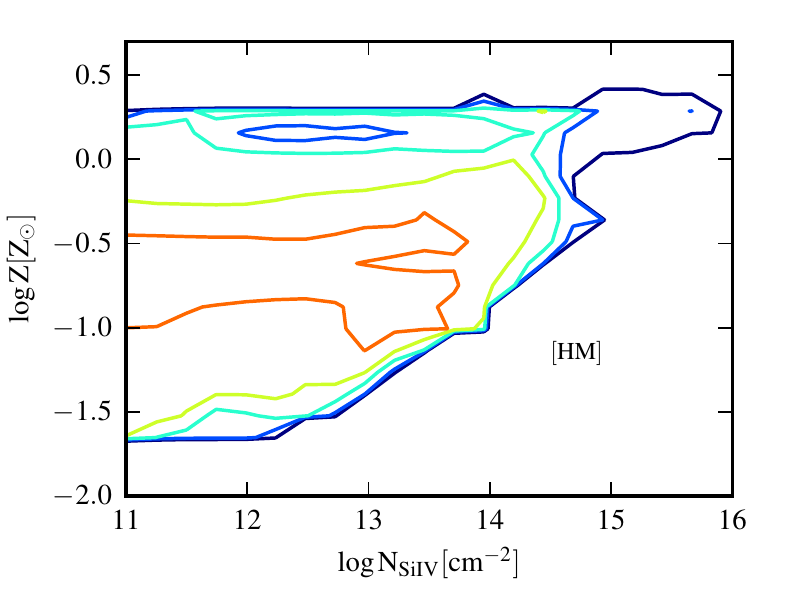}}
\hskip -0.0cm
\resizebox{3.4in}{!}{\includegraphics[angle=0]{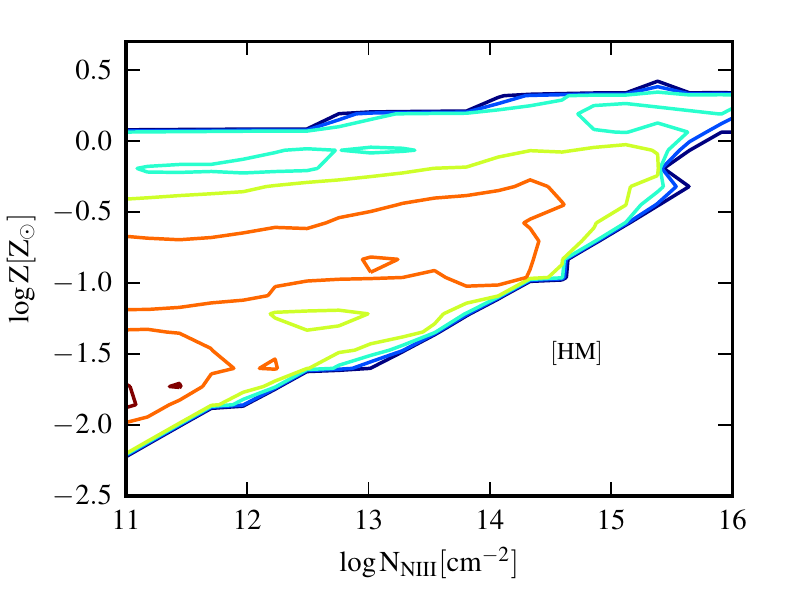}}
\vskip -0.0cm
\caption{
shows the number density of absorption lines
in the column density-metallicity plane 
for OVI (top-left panel),
NV (top-right panel),
NIII (bottom-right panel)
and
SiIV (bottom-left panel).
The contour levels are evenly spaced in log-scale spanning the range of 0.5 and half of maximum density in each panel. 
Only the HM case with varying N/H case is shown for all lines, because the difference between HM and HM+local cases
is found to be relatively small.
}
\label{fig:contour_ZN}
\end{figure}

\begin{figure}[!h]
\centering
\vskip -0.0cm
\hskip -1.0cm
\resizebox{3.4in}{!}{\includegraphics[angle=0]{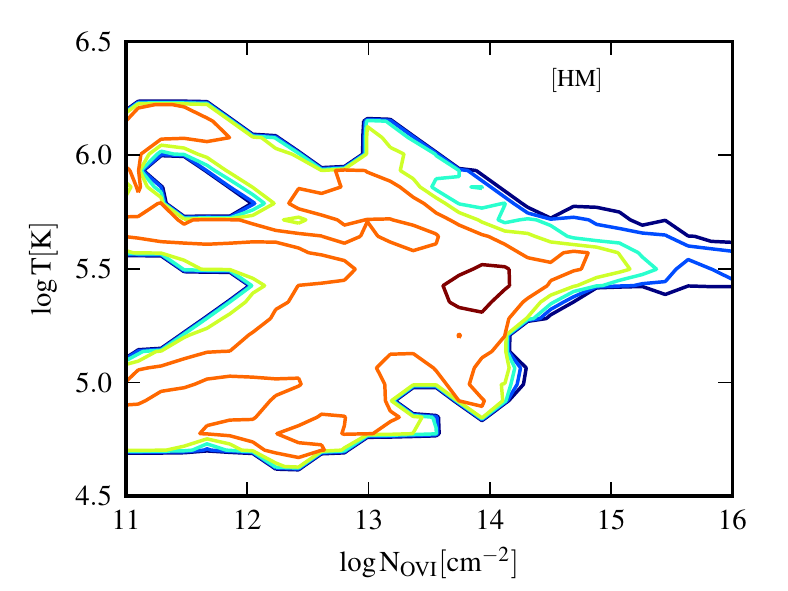}}
\hskip -0.0cm
\resizebox{3.4in}{!}{\includegraphics[angle=0]{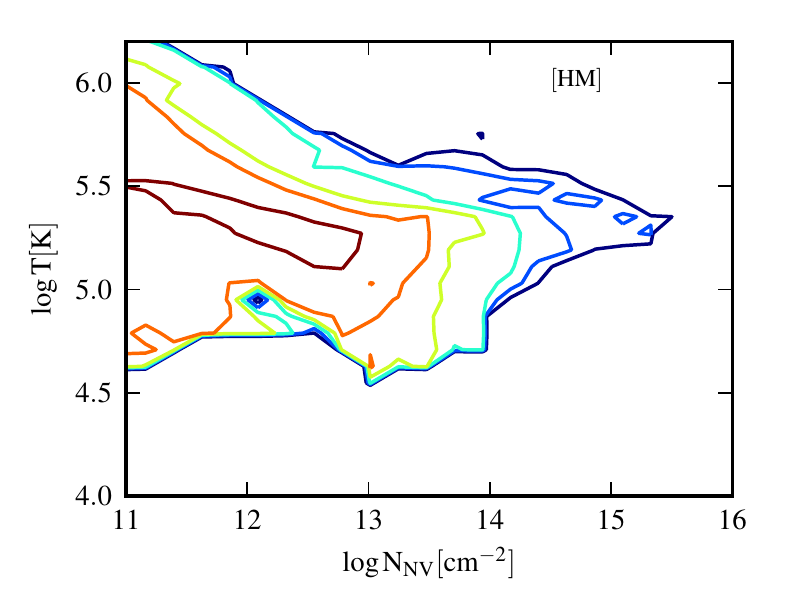}}
\vskip -0.0cm
\hskip -1.0cm
\resizebox{3.4in}{!}{\includegraphics[angle=0]{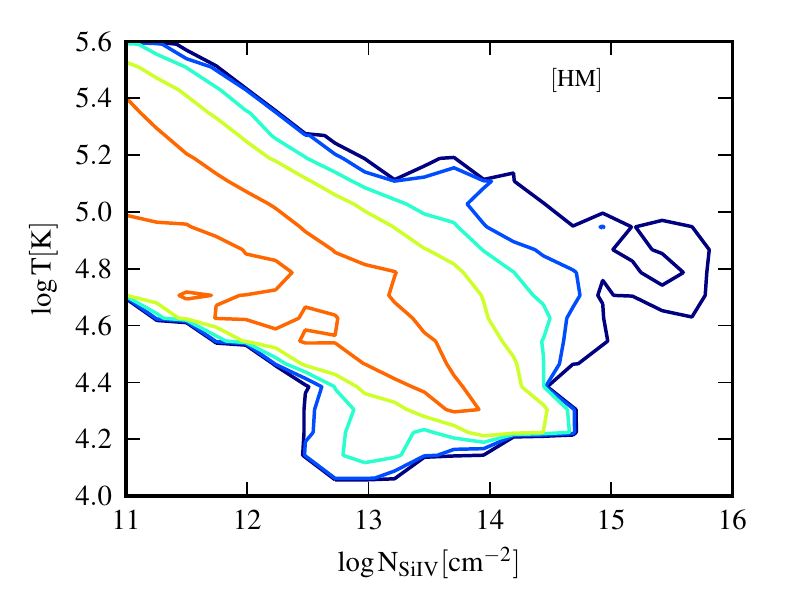}}
\hskip -0.0cm
\resizebox{3.4in}{!}{\includegraphics[angle=0]{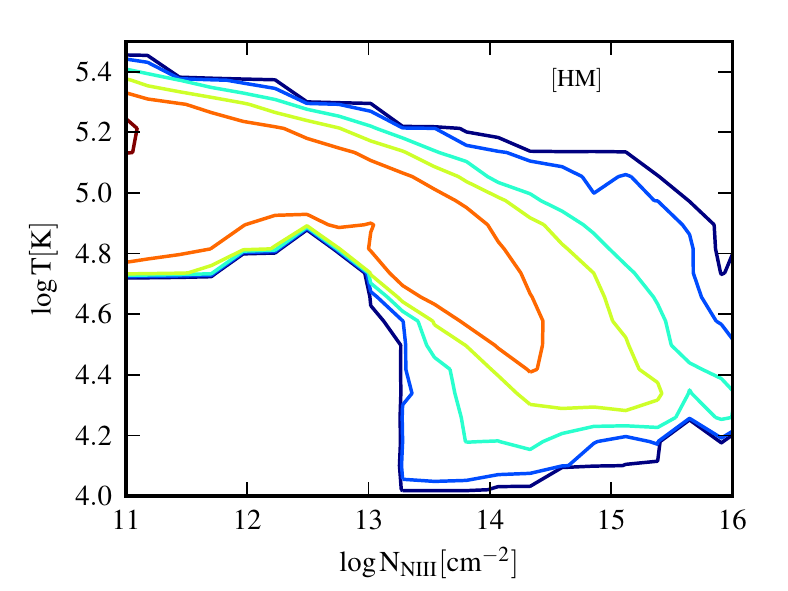}}
\vskip -0.0cm
\caption{
shows the number density of absorption lines
in the column density-temperature plane 
for OVI (top-left panel),
NV (top-right panel),
NIII (bottom-right panel)
and
SiIV (bottom-left panel).
The contour levels are evenly spaced in log-scale spanning the range of 0.5 and half of maximum density in each panel. 
Only the HM case with varying N/H case is shown for all lines, because the difference between HM and HM+local cases
is found to be relatively small.}
\label{fig:contour_TN}
\end{figure}

Finally, we turn to a closer examination of the physical conditions that give rise to 
the various absorption lines in our simulations shown above.
In Figure \ref{fig:contour_ZN} we 
show the number density of lines
in the column density-metallicity plane 
for OVI (top-left panel),
NV (top-right panel),
NIII (bottom-right panel)
and
SiIV (bottom-left panel).
Overall, while there is a significant span in metallicity, with as low as $\rm [Z/H]=-3.5$ at the low column density end for OVI, 
the vast majority of absorbers have metallicities falling into the range
$\rm [O/H]=-2$ to $-0.5$ for OVI,
$\rm [O/H]=-2$ to $-0.5$ for NV,
$\rm [O/H]=-2$ to $-0.5$ for NV
and
$\rm [O/H]=-1$ to $-0.5$ for SiIV.
At the high column density end, 
we see $\rm [O/H]\sim -0.5$ to $0$ for OVI,
$\rm [O/H]\sim -0.5$ to $0.5$ for NV,
$\rm [O/H]\sim 0$ to $0.5$ for NIII
and
$\rm [O/H]\sim 0$ to $0.5$ for SiIV.
These trends and significant disparities between different lines
are a results of complex multi-phase CGM with a very inhomogeneous metallicity distribution.
Simplistic collisional excitation/ionization models are unlikely to be able to 
capture all of the key elements of the physical processes involved and may lead to conclusions
that are not necessarily conformal to direct analyses of the simulations.

Figure \ref{fig:contour_TN}
shows the number density of lines
in the column density-temperature plane 
for OVI (top-left panel),
NV (top-right panel),
NIII (bottom-right panel)
and
SiIV (bottom-left panel).
To set the context of collisionally dominated ionization processes,
we note that, under the assumption of collisional ionization equilibrium, as in CLOUDY,
the peak temperature for the element in question with Half-Width-Half-Maximum 
is approximately 
$(3.0\pm 0.5)\times 10^5$K for OVI,  
$(2.0\pm 0.5)\times 10^5$K for NV,   
$7.5^{+5.0}_{-3.0}\times 10^4$K for NIII  
and 
$7.3^{+1.8}_{-2.2}\times 10^4$K for SiIV  
\citep[e.g.,][]{2007Gnat}.
A one-to-one comparison between each of these four peak temperature (and its width)
and the contour levels indicates that
for OVI and NV lines the collisional ionization dominates the process 
for creating OVI at $\rm N_{OVI}\ge 10^{14}$cm$^{-2}$ and 
NVI at $\rm N_{NV}\ge 10^{13}$cm$^{-2}$, respectively,
manifested in the horizontal extension of the contours pointing to the right at the temperature 
(with an appropriate width) in question. 
The same can be said about the SiIV line at the high column end
$\rm N_{SiIV}\ge 10^{15}$cm$^{-2}$;
however, at lower $\rm N_{SiIV}$ values ($\le 10^{15}$cm$^{-2}$),
the contours are no longer aligned horizontally, indicative of 
enhanced contribution of photoionization due to lower ionization potential of SiIII
($33.49$eV) versus say OV ($77.41$eV).
Similar statements about NIII lines to those for SiIV can be made due to similar reasons.
Overall, the similarity between OVI and NV lines suggests that collisional ionization processes
are dominant and results with respect to these two lines are relatively immune to 
uncertainties in the radiation field used.
However, the apparent insensitivity of results on the radiation field with detailed calculation we have performed, 
in the way of comparing HM and HM+local results,
indicates that the net effect due to an increase of radiation field 
is relatively small due to the large ranges of density and metallicity of gas involved,
although the actual situation appears to be more intertwined because of nonlinear relationships
between density, metallicity, ionization parameter and column density.
As an example, as shown earlier in Figure~\ref{fig:cover},
a stronger radiation field tend to increase the abundance of NIII,
although the difference between HM and HM+local cases for is apparently minor, 
seeming to suggest conflictingly that collisional processes are the primary powering source for producing NIII.
A more thorough theoretical study and a more detailed comparison to observations 
will be desirable, when a larger observational sample with more sensitive column density detection limits becomes available.
As we have demonstrated, a fraction of a dex increase in sensitivity may warrant
a revisit to a detailed comparison.

\section{Conclusions}

In light of a recent conclusion that the observed line ratios of UV absorbers
in the CGM may pose a significant challenge for theoretical models
\citep[][]{2016Werk},
we study five UV absorption lines,
OVI $\lambda \lambda$1032, 1038,
SiIV $\lambda \lambda$1394, 1402,
NV $\lambda \lambda$1239, 1243,
NIII $\lambda $990,
NII $\lambda $1084,
in the CGM of simulated galaxies,
utilizing {\it ab initio} ultra-high 
resolution ({\color{red}\bf LAOZI}) 
hydrodynamical simulations.
Our simulated galaxies are chosen to have 
stellar masses and star formation rates similar to their observed counterpart.

We examine uncertainties related to the radiation background by computing separately
for two cases of radiation field,
one with the HM radiation background and the other with both HM and local radiation due to hot gas
in the host galaxy.
Separately and orthogonally, we examine two separate cases of nitrogen to oxygen ratio,
in one case with constant N/O and in the other with varying N/O that is  
theoretically consistent with two different synthetic sources of nitrogen and
observationally confirmed by independent observations.

\citet[][]{2016Werk} find constant density photoionization models to be excluded by the data. They find
collisional (both in and out of equilibrium) ionization to be only broadly consistent with the data. They
suggest either collisionally-ionized gas cooling behind a fast shock or a highly structured 
gas photo-ionized by a local high energy source as plausible models to account of the observed 
OVI column density range and line ratios. In contrast, we do not find significant difficulty 
in accounting for the same observational data in our cosmological simulations that  
capture the complex multi-phase structure of the CGM,
as reflected by the acceptable KS test p-values for column density ratios of four pairs of lines 
NV/OVI, SiIV/OVI, NIII/OVI and NII/OVI examined.

Inter-comparisons between results from different models employing different radiation fields
in our simulations and comparisons between properties of the absorbers and expectations
of collisional ionization indicate that collisional ionization play a major role
in producing all the lines studied in realistic CGM produced in cosmological simulations.
Photoionization process plays a significant role as well, to a varying degree, depending
on the ion in question, although it seems clear that for NV and OVI lines photoionization effect is relatively minor.

The success of our largely collisional ionization model in all cases is, however,
in a significant part, due to very accommodative observational line ratio data points 
that are dominated, in number, by upper limits rather than actual detections.
We find the apparent satisfactory agreement between simulations and extant observational data can be strongly tested
and different cases (different radiation fields and different N/O ratio assumptions) differentiated, 
if most of the current upper limits in the observational data become detections.
To demonstrate the power, we show, as an example,
that if the upper limits of NV/OVI become detections with values that are lower by a mere $\sim 0.20$ dex
than their respective current upper limits, 
the KS p-value for the NV/OVI line ratio becomes $\sim 0.01$ for the constant N/O case 
and $\sim 0$ for the varying N/O case. 
If, on the other hand, the actual detection values turn out to be lower than current upper limits by 
$0.6-0.7$ dex, 
then the varying N/O case obtains a satisfactory p-value of $(0.5,0.8)$,
whereas the
constant N/O case is endowed with a p-value equal to zero.
Thus, it is highly desirable to increase the observational sensitivity and/or enlarge observational
data sample size, in order to have a definitive test.

\vskip 1cm

We are grateful to Jessica Werk for sharing observational data with us prior to publication and stimulating
discussion.
We thank J. Xavier Prochaska for useful discussion.
We have used the very useful and versatile 
analysis software yt version 2.6 \citep{2011Turk}
for some of our analysis.
Computing resources were in part provided by the NASA High-
End Computing (HEC) Program through the NASA Advanced
Supercomputing (NAS) Division at Ames Research Center.
The research is supported in part by NASA grant NNX11AI23G.


\end{document}